\documentclass[reprint,aps,pra]{revtex4-2}
\usepackage{packages}
\begin{document}
\emergencystretch=3em
\hbadness=10000
\preprint{APS/123-QED}

\title{A Variational Quantum Algorithm for Multi-Skill Project Scheduling}

\author{Raul Martinez}
    \affiliation{Quantum Technologies Department, ITECAM}
\author{Miguel Sánchez-Beato}
    \affiliation{Quantum Technologies Department, ITECAM}
\author{Matilde Osa}
    \affiliation{Quantum Technologies Department, ITECAM}
\author{Jorge Parra}
    \affiliation{Quantum Technologies Department, ITECAM}
\author{Mario Calonge}
    \affiliation{Quantum Technologies Department, ITECAM}
\date{\today}

\begin{abstract}
\begin{singlespace}
Job scheduling is an NP-hard optimization problem with clear industrial relevance. In practice, distributing work across a team is not only about meeting deadlines or improving productivity; it also affects fatigue, stress, and employee well-being. Variational Quantum Algorithms (VQAs) provide a promising way to address these problems through QUBO formulations, although their scalability on near-term NISQ hardware is still limited by Hilbert-space growth and barren plateaus. In this work, we simulate a VQA for a realistic Multi-Skill Resource-Constrained Project Scheduling Problem (MS-RCPSP), including worker capacities, skill matrices, task priorities, deadlines, and calendar stability during rescheduling. To keep the problem within a near-term circuit size, we use a temporal batch decomposition that solves one day at a time, propagates the remaining workload, and combines it with a logarithmic binary encoding that keeps every batch at or below 30 qubits. Because the ground state of a QUBO Hamiltonian is a computational basis state, entanglement is not required to represent the solution, raising the question of whether it helps or hinders ansatz training. We address this question by comparing the linear and brick-wall entangling topologies of a hardware-efficient ansatz with an entanglement-free reference that has the same number of parameters and by training every configuration with COBYLA and with Adam using SPSA gradient estimation. For the instance and initialization studied, the entanglement-free reference produced feasible schedules under both optimizers. With COBYLA, it produced the feasible schedule closest to the required workload at a fraction of the simulation cost of the entangled topologies, indicating that entanglement did not improve the solutions found by the VQA.

\end{singlespace}

\end{abstract}

\maketitle
\onehalfspacing

%

\section{Introduction} \label{sec:Introduction}
Effective resource allocation and job scheduling are central challenges in modern enterprise management. Distributing workloads among a finite set of workers is not only a matter of meeting tight deadlines; it also plays an important role in maintaining team well-being and reducing fatigue and psychological stress~\cite{bowling2015metaanalytic}. Mathematically, scheduling problems belong to the family of NP-hard combinatorial optimization problems. The \textbf{MS-RCPSP} (Multi-Skill Resource-Constrained Project Scheduling Problem)~\cite{hartmann2010survey}~\cite{bellenguez2005lower} generalizes the classical job scheduling problem by introducing two industrially relevant limitations: each worker has a heterogeneous skill set that allows them to work on only some jobs, together with a finite daily capacity that must be respected throughout the planning horizon. As the number of tasks and scheduling days grow, the solution space expands exponentially, making exact classical solvers computationally intractable for realistic instances and forcing practitioners to rely on heuristics that may reduce solution quality~\cite{quoc2023meminv}~\cite{myszkowski2015hybrid}.

Quantum computing has emerged as a computational paradigm that may help address these types of problems~\cite{preskill2018nisq}. A common approach maps the classical constraints of the problem into a \textbf{QUBO} (Quadratic Unconstrained Binary Optimization) formulation, in which the desired optimal solution corresponds to the ground state of an equivalent Ising Hamiltonian~\cite{glover2019tutorial}. Several quantum approaches can then be used to search for this minimum-energy state. Along with quantum annealing, which is suited to analog devices~\cite{kadowaki1998quantum}~\cite{venturelli2015quantum}, \textbf{VQAs} (Variational Quantum Algorithms) have become a leading strategy for gate-based quantum processors (QPUs) in the current \textbf{NISQ} (Noisy Intermediate-Scale Quantum) era~\cite{cerezo2021variational}. VQAs delegate the optimization to a parameterized quantum circuit called the \textbf{ansatz}, whose parameters are iteratively updated by a classical optimizer to minimize the expectation value of the problem Hamiltonian~\cite{Benedetti_2019}.

Despite their appeal, VQAs face serious algorithmic bottlenecks as the qubit count grows. Both the number of parameters and the circuit depth scale with the system size, creating constraints that directly affect ansatz training. As the number of trainable parameters increases, the optimization can encounter \textbf{barren plateaus}, regions of the optimization landscape where gradients vanish~\cite{mcclean2018barren}. This prevents classical optimizers from updating the circuit effectively and making meaningful progress. Furthermore, the limited connectivity and coherence times of NISQ devices severely restrict the depth of variational circuits, reducing circuit expressivity~\cite{sim2019expressibility} and, ultimately, constraining the quality of the outcome. Consequently, studying the behavior of VQAs at industrially meaningful scales currently relies on HPC architectures that can faithfully simulate the entire circuit without introducing hardware noise.

In this work, we present a quantum simulation of a realistic enterprise job-scheduling problem through a complete VQA pipeline. Using the matrix product state simulator of a Bull Qaptiva 802 appliance, whose programming interface is provided by the myQLM software stack~\cite{myqlm}, we evaluate the performance of VQAs for solving an MS-RCPSP instance. Our QUBO formulation explicitly models workers' daily capacities, individual skills, task priorities, deadlines, and a calendar-stability term that penalizes deviations from a previous schedule whenever the algorithm is invoked in rescheduling mode. To keep the circuits within a trainable size, we introduce a temporal batch decomposition that solves a single day at a time and propagates the remaining workload to subsequent batches, together with a logarithmic binary encoding. This approach keeps every per-batch QUBO at or below 30 binary variables, and therefore at or below 240 trainable parameters, while preserving the full multi-day structure of the problem.

The central question of this paper is whether entanglement contributes to the trainability of a hardware-efficient ansatz on a realistic scheduling QUBO. Because the ground state of an Ising Hamiltonian derived from a QUBO is a computational basis state, a single layer of single-qubit rotations is already expressive enough to represent the optimal solution. Previous work has examined how entanglement relates to problem hardness in quantum variational optimization~\cite{diezvalle2021entanglement}. We therefore investigate whether entangling layers help the optimizer reach that solution or instead make the landscape harder to navigate. We compare three ansatz families that share the same rotation layers and differ only in their entangling stage: a linear topology, in which each qubit is entangled with its successor; a brick-wall topology, in which two-qubit gates are arranged in alternating even-odd layers; and an entanglement-free reference, in which the entangling stage is removed. All three families use $r = 3$ repetitions and therefore have the same number of trainable parameters, so differences between them can be attributed to the entangling stage rather than to the size of the parameter space. As a second design axis, we train every ansatz with COBYLA and with Adam using SPSA gradient estimation, giving six ansatz-optimizer configurations. All six start from the same randomly drawn initial parameter vector, so the comparison between them is paired with respect to initialization.

%

\section{Problem Formulation} \label{sec:QUBO}
\subsection{Problem statement}
We consider an instance of the MS-RCPSP defined over a finite planning horizon of $D$ working days~\cite{bellenguez2007branch}~\cite{hartmann2010survey}. The instance includes a complete set of $N_W$ workers, defined in Eq.~\ref{eq:workers}, together with a set of $N_T$ tasks, defined in Eq.~\ref{eq:tasks}, that must be completed before the deadlines defined in Eq.~\ref{eq:deadlines}. The schedule must satisfy two restrictions. First, every worker has a daily capacity, measured in working hours, that corresponds to their individual work shift. Second, each worker $i$ has an individual skill profile $\mathcal{S}_i \subseteq \mathcal{T} = \{1, \dots, N_T\}$ that identifies the tasks they are qualified to perform. Thus, a worker cannot be assigned hours for a task that they are not qualified to perform.

In addition, each task $j$ is characterized by a total workload $W_j$, measured in working hours, that must be completed before its deadline $d_j \in \{1, \dots, D\}$, and by a priority weight $\pi_j$ that reflects its relative urgency within the overall schedule. A valid schedule is an assignment of working hours that respects every worker's daily capacity and skill profile and, in turn, completes the total workload of every task while meeting all task deadlines.
\begin{subequations}
    \begin{equation}\label{eq:workers}
    \mathcal{W} = \{1, \dots, N_W\}
\end{equation}
\begin{equation}\label{eq:tasks}
    \mathcal{T} = \{1, \dots, N_T\}
\end{equation}
\begin{equation}\label{eq:deadlines}
    \mathcal{D} = \{1, \dots, D\}
\end{equation}
\end{subequations}

\subsection{Binary encoding}
Mapping the MS-RCPSP directly to a QUBO requires choosing a discrete representation for the number of working hours assigned to each \textbf{(worker, task, day)} tuple. To reduce the number of variables in the final QUBO and, consequently, the number of qubits required to construct our VQA circuit, we adopt a \textbf{logarithmic binary encoding} of the assigned working hours. Time is discretized into slots of fixed duration $h$; in our experiments, we choose $h=2$ hours. With this discretization, a worker's daily capacity, $C_i$, can be expressed as $C_i' = \lfloor C_i/h\rfloor$ integer slots. We introduce the binary variables shown in Eq.~\ref{eq:binary}, indexed by an exponent $p\in \{0, 1, \dots, P_i\}$, where $P_i = \lfloor \log_2(C_i/h)\rfloor$.

\begin{equation}\label{eq:binary}
    x_{i, j, p, d} = \begin{cases}
        1 \: \text{if worker $i$ is assigned $h\cdot2^p$}\\\:\:\: \text{hours to task $j$ on day $d$}\\ 0 \: \text{otherwise}
    \end{cases}
\end{equation}

This encoding greatly reduces the number of variables. A worker with an 8-hour shift would require 8 binary variables per task under a one-hot encoding. By discretizing time into 2-hour windows, we reduce this number to 4; by introducing logarithmic binary encoding, we further reduce the number of variables required per worker, task, and day to 3. The number of hours assigned to any worker $i$ on any day $d$ can be expressed as shown in Eq.~\ref{eq:slots}.

\begin{equation}\label{eq:slots}
n_{i, d} = \sum_{j\in \mathcal{S}_i}\sum_{p=0}^{P_i}h\cdot2^px_{i, j, p, d}
\end{equation}

This binary encoding is not, by itself, capacity-tight, since it is possible to represent an hour assignment that exceeds a worker's daily capacity. We later suppress such violations through QUBO penalty terms~\cite{montanez2024unbalanced}. In addition, only feasible variables are created. This means that any variable $x_{i, j, p, d}$ for which $j\notin \mathcal{S}_i$ or $d > d_j$ holds is not introduced in the QUBO formulation. This pre-filtering reduces the number of qubits required to encode an instance without altering the optimal solution.

\subsection{QUBO formulation}
The MS-RCPSP is mapped to a QUBO model by encoding each requirement as a quadratic penalty term whose minimum corresponds to the optimal solution of the system~\cite{lucas2014ising}. The resulting QUBO takes the standard form shown in Eq.~\ref{eq:QUBO} and is assembled from an objective term and three constraint terms, each weighted by a positive Lagrange multiplier according to its relative importance.

\begin{equation}\label{eq:QUBO}
    \mathcal{H}_{QUBO} = \sum_{a}Q_{aa}x_a + \sum_{a<b} Q_{ab}x_ax_b
\end{equation}

\subsubsection{Objective function}
The objective term in Eq.~\ref{eq:objective} expresses a preference for schedules that complete each task as early as possible. Each variable $x_{i, j, p, d}$ contributes a cost proportional to $d$, the index of the day on which the assignment is placed:

\begin{equation}\label{eq:objective}
    \mathcal{H}_{obj} = \sum_{i, j, p, d\leq d_j} d \cdot x_{i, j, p, d}
\end{equation}

Intuitively, this term encourages the solver to fill earlier days first. Completing the same job on day $d=1$ introduces a smaller penalty than completing it on day $d=5$. It is important to note that the effect of this term is only relevant in instances in which we can optimize several days within the same QUBO. In our experiments, we optimize a single day at a time, so $d$ is constant within each batch and this term reduces to a uniform linear penalty on every assignment, weighted by a small multiplier $\lambda_{\text{obj}}$, which does not encode any preference between days.

\subsubsection{Constraint 1: Worker daily capacity}
A valid solution must respect the daily capacity $C_i$ of every worker $i$ on each scheduled day $d$. The total hours assigned to a worker, $n_{i,d}$, must respect that capacity. This restriction is introduced as a quadratic penalty $\mathcal{H}_{cap}$:

\begin{equation}\label{eq:capacity}
    \mathcal{H}_{cap} = \sum_{i, d} \left(\sum_{j\in\mathcal{S}_i}\sum_ph\cdot 2^p \cdot x_{i, j, p, d} - C_i\right)^2
\end{equation}

Enforcing the inequality $n_{i, d} \le C_i$ exactly would require additional slack variables and, therefore, additional qubits. We instead use the equality penalty of Eq.~\ref{eq:capacity}, which penalizes both over- and under-utilization and drives every active worker toward full use of their daily capacity. Capacity violations reported in Section~\ref{sec:Results} refer only to assignments with $n_{i, d} > C_i$. We set the Lagrange multiplier of $H_{cap}$, $\lambda_{cap}$, sufficiently large because worker capacity takes priority over the other scheduling requirements, so that capacity violations are too costly for any other term to compensate.

\subsubsection{Constraint 2: Task hour targets}\label{sec:Target}
Each task $j$ must accumulate a total of $W_j$ working hours across all qualified workers and all days up to its deadline $d_j$. In our batch decomposition strategy, discussed in more detail below, we break this global restriction into a per-batch requirement that distributes the workload evenly over the days remaining until the deadline. The global constraint is introduced, as shown in Eq.\ref{eq:target_global}, as a quadratic penalty.

\begin{equation}\label{eq:target_global}
    \mathcal{H}_{\mathrm{tgt}}^{\mathrm{global}} = \sum_{j} \left(\sum_{\substack{i\,:\,j\in\mathcal{S}_i}}\sum_{p,\,d \leq d_j}h\cdot 2^p\cdot x_{i, j, p, d} - W_j\right)^2
\end{equation}
The batch decomposition strategy used in our experiments replaces this global requirement with a per-batch one. For each batch $b$, we compute a proportional task-hour target $T_j^{(b)}$ that distributes the remaining workload for task $j$ evenly over the days remaining until its deadline. Solving the QUBO independently for each batch introduces a new issue. Because each batch optimizes its own targets, the solver has no incentive to favor urgent tasks over less urgent ones. In early batches, working hours may be assigned to tasks with looser deadlines, leaving tight-deadline tasks unfinished by the time their deadlines arrive, even when the total worker capacity would have been sufficient to complete them all.

To preserve this global priority order at the local level, we introduce a per-task priority weight $\pi_j$, which represents the urgency of each task. Multiplying every target term by $\pi_j$ biases the Hamiltonian ground state toward allocating working hours to high-priority tasks first while still penalizing under-allocation and over-allocation around the target. The resulting per-batch target term is given in Eq.\ref{eq:target_batch}.

\begin{equation}\label{eq:target_batch}
    \mathcal{H}_{\mathrm{tgt}} = \sum_{j} \pi_j\left(\sum_{\substack{i:j\in\mathcal{S}_i}}\sum_{p,d \leq d_j}h\cdot 2^p\cdot x_{i, j, p, d} - T_j^{(b)}\right)^2
\end{equation}

\subsubsection{Constraint 3: Task-switching penalty}
Operational regularity is a desirable property of any practical schedule: a worker who focuses on a single task incurs fewer context-switching costs than one who alternates between several tasks on the same day. We model this preference with a quadratic penalty that activates whenever two variables associated with the same worker and day but different tasks are simultaneously set to one:

\begin{equation}\label{eq:switching}
    \mathcal{H}_{sw} = \sum_{i, d}\sum_{p_1, p_2}\sum_{j_1, j_2 \in \mathcal{S}_i/ j_1 < j_2} x_{i, j_1, p_1, d}\cdot x_{i, j_2, p_2, d}
\end{equation}

\subsubsection{Final QUBO}
Combining the constraints and the objective function gives the final QUBO in Eq.\ref{eq:QUBO_final}.

\begin{equation}\label{eq:QUBO_final}
    \mathcal{H}_{QUBO} = \lambda_{obj}\mathcal{H}_{obj} + \lambda_{cap}\mathcal{H}_{cap} + \lambda_{tgt}\mathcal{H}_{tgt} + \lambda_{sw}\mathcal{H}_{sw}
\end{equation}

We choose the Lagrange multipliers according to the operational hierarchy of the problem, which gives the order shown in Eq.~\ref{eq:order}. Capacity violations are forbidden in practice and receive the largest penalty. Task-target and switching penalties are not hard constraints; thus, they receive lower penalty multipliers. Finally, the objective function does not encode any preference between days in single-day batches and receives the smallest multiplier.

\begin{equation}\label{eq:order}
    \lambda_{cap}>\lambda_{sw}\geq\lambda_{tgt}>\lambda_{obj}
\end{equation}

The Lagrange multiplier values used in the experiments are listed in Table~\ref{table:lagrange}.

\begin{table}[htbp]
\centering
\begin{tabular}{cc | cc}
 Lagrange multiplier&&& Value \\
\hline
$\lambda_{cap}$ &&& 400 \\
$\lambda_{tgt}$ &&& 0.02 \\
$\lambda_{sw}$ &&& 20 \\
$\lambda_{obj}$ &&& 0.01 \\
\end{tabular}
\caption{Lagrange multiplier values used in the experiments.}
\label{table:lagrange}
\end{table}

\subsection{Rescheduling instances}
In industrial planning, a schedule rarely remains unchanged once it has been computed. New tasks may arrive, deadlines may be modified, and workers might become temporarily unavailable. In all these cases, a useful planner should be able to produce an updated schedule that incorporates the new information without discarding the structure of the previous one~\cite{herroelen2005project}. Workers also expect predictable plans, so large unnecessary reshuffles reduce the practical value of the optimization. We refer to this property as calendar stability, and we encode it as an additional optional QUBO term that is activated whenever the algorithm is invoked in rescheduling mode.

Let $x_{old}$ denote the value that a given variable $x$ takes in the previously computed schedule. For each variable that exists in both the previous and current instances, we define the stability term as a quadratic penalty on the deviation between the current value and the reference value. We refer to the set of variables that are common to both the previous and the current instances as $\mathcal{X}$.

\begin{equation}\label{eq:stability}
    \mathcal{H}_{stab} = \sum_{(i,j,p,d) \in \mathcal{X}}(x_{current}-x_{old})^2
\end{equation}

Expanding this quadratic penalty shows that it can be reduced to a purely linear contribution, as shown in Eq.\ref{eq:stability_2} and Eq.\ref{eq:stability_3}.

\begin{equation}\label{eq:stability_2}
    \mathcal{H}_{stab} = \sum_{(i,j,p,d) \in \mathcal{X}}( x_{current}^2 - 2\cdot x_{current}\cdot x_{old} + x_{old}^2)
\end{equation}
\begin{equation}\label{eq:stability_3}
    \mathcal{H}_{stab} =  \sum_{(i,j,p,d) \in \mathcal{X}}(1 - 2\cdot x_{old})x_{current}
\end{equation}
The simplification from Eq.\ref{eq:stability_2} to Eq.\ref{eq:stability_3} follows from removing the term $(x_{old})^2$ from the optimization landscape and using the Boolean relation $x_{current}^2 = x_{current}$.

The linearity of this term has a practical consequence: it does not contribute any quadratic coupling to the QUBO matrix and therefore does not increase the density of the Ising Hamiltonian's interaction graph.

The penalty multiplier $\lambda_{stab}$ should remain low enough to let the optimizer freely reschedule work assignments when needed, while still being high enough to maintain calendar stability and minimize the changes required to incorporate new information into the calendar.

The stability term is part of the formulation but is not activated in the experiments reported here.
\subsection{Conversion to Ising Hamiltonian}
The QUBO formulation introduced previously is defined over binary variables $x \in \{0,1\}$. To make it suitable for a VQA, it must be rewritten as the expectation value of a Hermitian operator acting on $n$ qubits, where $n$ equals the number of binary variables in the QUBO. This is achieved through the canonical change shown in Eq.\ref{eq:ising}, which maps each binary variable $x_a$ to a Pauli-$Z$ operator~\cite{lucas2014ising}.

\begin{equation}\label{eq:ising}
    x_a = \frac{I - Z_a}{2}
\end{equation}

Here, $Z_a$ denotes the Pauli-$Z$ operator acting on qubit $a$, and $I$ is the identity operator. Under this substitution, $x_a$ takes the value $0$ when the qubit is in the $+1$ eigenstate of $Z_a$ and the value $1$ when the qubit is in the $-1$ eigenstate, that is, in the computational basis states $|0\rangle$ and $|1\rangle$, respectively.

After applying this change, the numerical scale of the coefficients of the operators composing the Ising Hamiltonian matters for variational optimization. Normalizing the coefficients of the Pauli terms has been reported to improve the feasibility of sampled solutions in QAOA-based optimization of constrained routing problems~\cite{azfar2026quantumassistedvehicleroutingrealizing}. Following this practice, we normalize the Hamiltonian as shown in Eq.~\ref{eq:normalization}, using the root mean square of the coefficients as the scale factor.

\begin{equation}\label{eq:normalization}
    \rho = \sqrt{\frac{1}{M}\sum_{k=1}^{M}\left|c_k\right|^2} \qquad
    \mathcal{H}_{\mathrm{Ising}}^{\mathrm{norm}} = \frac{\mathcal{H}_{\mathrm{Ising}}}{\rho}
\end{equation}

Here, $M$ is the number of non-identity Pauli terms, while $c_k$ denotes the coefficient of the $k$-th term. This rescaling preserves the location of the ground state of the Ising Hamiltonian while bringing the magnitude of the cost function into a comparable range across batches of different sizes.

%

\section{Circuit Construction and Simulations} \label{sec:Development}
\subsection{Hardware-efficient ansatz}
The QUBO formulation above reduces the MS-RCPSP to the search for the ground state of an Ising Hamiltonian. We implement this ground-state search as a VQA executed on a matrix product state (MPS) simulator. The VQA training pipeline repeatedly evaluates the expectation value of the Ising Hamiltonian on a parameterized circuit, called the ansatz, and updates its parameters to drive the state toward low-energy configurations.

The ansatz used in this work belongs to the family of hardware-efficient ansatz circuits~\cite{zhang2023single, nakaji2021expressibility}, in which expressivity is built up by alternating layers of parameterized single-qubit rotations with layers of fixed two-qubit entangling gates~\cite{sim2019expressibility}. For QUBO problems on gate-based devices, the canonical variational approach is QAOA (Quantum Approximate Optimization Algorithm)~\cite{farhi2014qaoa}, whose ansatz alternates between a problem Hamiltonian and a mixing Hamiltonian with a structure derived directly from the cost function. However, a hardware-efficient ansatz allows us to design the entangling topology freely, enabling a comparison of different entangling stages, including their complete removal, on the same QUBO instances. These ansatz circuits are well suited to NISQ-era execution because their native two-qubit gates and shallow structure keep them compatible with the limitations of current hardware~\cite{preskill2018nisq}. For this work, in which all experiments are executed in simulation, the hardware-efficient ansatz and its two-qubit gate geometry provide a clear basis for comparing different entanglement topologies.

The circuit acts on $n$ qubits, where $n$ equals the number of binary variables in the QUBO instance of the current batch. Qubits are indexed by iterating over workers, then over tasks, and finally over the exponents $p$ of the logarithmic encoding, so the variables of each worker occupy consecutive positions in the register. A single repetition of the ansatz can be divided into two stages. The first stage independently applies an $R_x$ rotation followed by an $R_y$ rotation to every qubit, with a separate trainable parameter for each rotation. The second stage applies a layer of $CNOT$ gates whose arrangement defines the entanglement topology. The complete ansatz consists of $r$ consecutive repetitions of $R_x-R_y-CNOT$ layers and ends with a final layer of $R_x-R_y$ rotations.

With this ansatz definition, the number of trainable parameters is $2n(r+1)$. In our experiments, we use $r = 3$ repetitions for all ansatz families, which corresponds to $240$ trainable parameters for the largest batches of $n = 30$ qubits.

Figures~\ref{fig:topologies_brickwall} and~\ref{fig:topologies_lineal} show the two entangled topologies under study. The two trainable circuits differ only in the arrangement of the $CNOT$ gates within an entangling layer.

\begin{center}
\begin{minipage}{0.95\columnwidth}
    \centering
    \includegraphics[width=\linewidth]{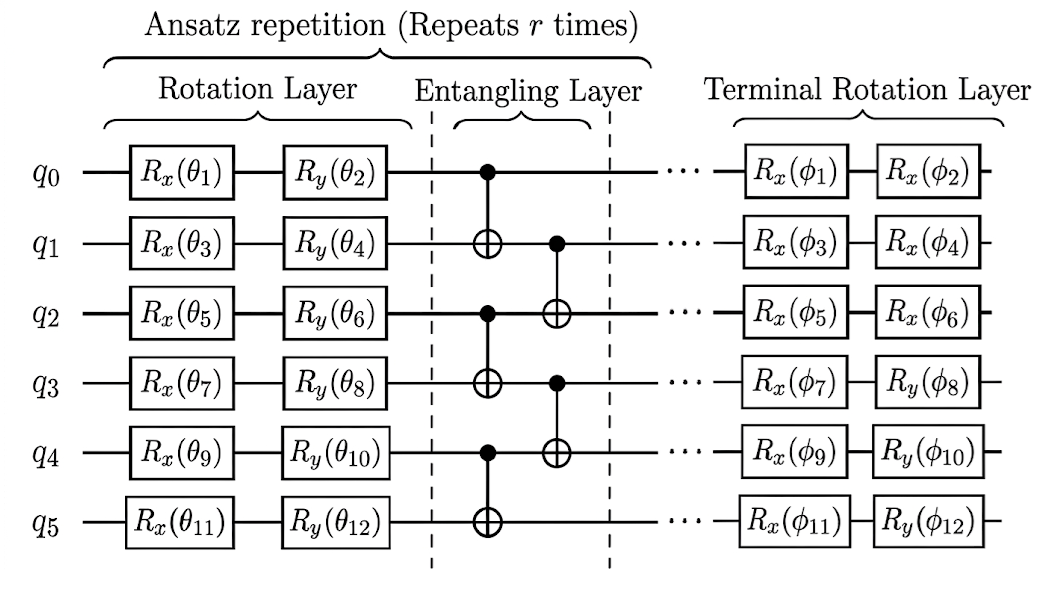}
    \makeatletter
    \def\@captype{figure}
    \makeatother
    \caption{Brick-wall entanglement topology of the ansatz.}
    \label{fig:topologies_brickwall}
\end{minipage}
\end{center}
\begin{center}
\begin{minipage}{0.95\columnwidth}
    \centering
    \includegraphics[width=\linewidth]{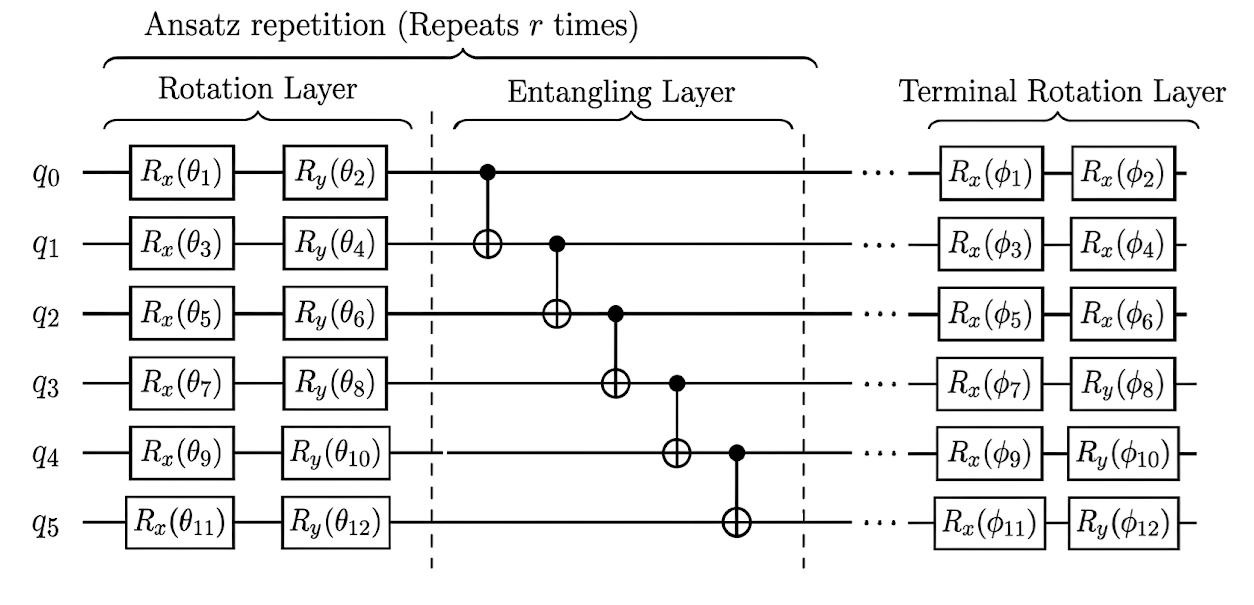}
    \makeatletter
    \def\@captype{figure}
    \makeatother
    \caption{Linear entanglement topology of the ansatz.}
    \label{fig:topologies_lineal}
\end{minipage}
\end{center}

In the \textbf{brick-wall} topology, the entangling layer can be divided into two sub-layers. The first applies $CNOT$ gates to even-indexed pairs, such as (0, 1), (2, 3), and so on. The second sub-layer applies $CNOT$ gates to odd-indexed pairs of consecutive qubits, such as (1, 2), (3, 4), and so on.

In the \textbf{linear} topology, the entangling layer consists of a single sequence of $CNOT$ gates applied to all consecutive qubit pairs, such as (0, 1), (1, 2), and so on. This topology has one advantage and one disadvantage: the cumulative connectivity reached at the end of a single entangling layer is larger than that of brick-wall, since information can propagate from one end of the qubit chain to the other in a single sweep. However, the gates cannot be parallelized, increasing the effective circuit depth and, therefore, making the circuit harder to execute on current QPUs due to the difficulty of maintaining qubit coherence over time.

The two topologies differ in circuit depth and connectivity. The brick-wall topology preserves a constant-depth entangling layer at the cost of slower information propagation. In contrast, the linear topology achieves long-range correlations within a single layer at the cost of a depth that grows proportionally to the qubit count. Comparing them on the same QUBO instances and under the same training procedure allows us to evaluate their impact on training behavior and final solution quality.

In the entanglement-free reference, the entangling stage is removed. Consecutive rotations acting on the same qubit then compose into a single SU(2) rotation, so the state remains a product state for any $r$. This family has the same expressivity as a single rotation layer while carrying exactly the same number of trainable parameters as the entangled families.

The linear and brick-wall topologies apply CNOT gates to exactly the same pairs of neighboring qubits, $(0, 1), (1, 2), \ldots, (n-2, n-1)$, and therefore contain the same number of CNOT gates, $n-1$, per entangling stage. They differ only in the order in which these gates are applied. Because CNOT gates that share a qubit do not commute, these different orders produce different unitaries. This design separates two effects: the effect of entanglement, measured by comparing the entangled topologies with the entanglement-free reference, and the effect of the entangling order, measured by comparing the linear and brick-wall topologies.

Since the ground state of the Ising Hamiltonian is a computational basis state, a single layer of rotations is already expressive enough to represent the optimal schedule. The question addressed in this work is therefore not whether the entangled topologies can represent the solution, but whether entanglement helps or hinders the optimizer in finding it.

\subsection{Batch decomposition strategy}\label{sec:batch}
Encoding the full 12-day horizon of our instance as a single QUBO would require 258 binary variables and, with $r=3$, approximately $2064$ trainable parameters, a regime in which training a variational circuit becomes impractical. We therefore adopt a temporal decomposition that divides the planning horizon of $D$ days into $B$ batches of $d_{win}$ consecutive days. We set $d_{win} = 1$ in all our simulations, which minimizes the qubit count by solving each batch as an independent QUBO. The per-batch target $T_j^{(b)}$ introduced in Section~\ref{sec:Target} is computed by distributing the remaining workload of task $j$ evenly over the days still available up to its deadline. After each batch is solved, the remaining workloads are updated and propagated to the next batch. Because each batch is solved independently, the solver has no visibility of future batches. For this reason, we introduce the priority-weight term discussed in Section~\ref{sec:Target}, which preserves the global ordering of urgency at the local level. Multi-day batches trade QUBO size for global coherence and remain a straightforward extension. In this work, we operate at the tightest extreme of this trade-off. With $d_{win} = 1$, the objective term of Eq.~(\ref{eq:objective}) reduces to a uniform linear term within each batch and does not encode any preference between days. For the instance considered in this work, the batch size decreases from 30 qubits in the first few days to 24, 19, and 13 qubits as tasks are completed or reach their deadlines. Thus, the exact days on which the size changes depend on the schedule produced by each configuration. Because each batch depends on the assignments made in the previous batches, only the first batch is an identical QUBO instance for all configurations; subsequent batches differ according to the schedule produced so far.

\subsection{Variational training pipeline}\label{sec:VQA}
For every batch in the planning horizon, the VQA executes a training loop that produces a candidate schedule for that batch's QUBO instance. The loop alternates between the quantum side, which runs the parameterized circuit and evaluates the expectation value of the Ising Hamiltonian for the resulting state, and the classical side, which updates the parameters of the ansatz based on those evaluations. Below, we describe the pipeline for a single batch. The same procedure is then repeated for every batch in the planning horizon.

The pipeline starts with the construction of the per-batch QUBO described in Section~\ref{sec:QUBO} and its conversion to the normalized Ising Hamiltonian. The number of binary variables in the QUBO determines the number of qubits $n$ in the parameterized circuit and, consequently, the number of trainable parameters in the variational loop. The initial values of these parameters are drawn from a Gaussian distribution with a standard deviation of $0.1$~\cite{zhang2022escaping}. When possible, the initial parameters are obtained through a \textbf{warm-start} mechanism, which reuses the optimized parameters of the previous batch whenever the two batches share the same number of trainable parameters~\cite{truger2024warmstarting}. When the QUBO size changes as tasks are completed or reach their deadlines, the warm-start is discarded, and the parameters are redrawn. For a given random seed, the initial parameter vector of the first batch is identical for all six configurations because all ansatz families share the same number of parameters.

Once the parameters are initialized, the training loop iterates over the cost function, defined as the expectation value of the Ising Hamiltonian. At every iteration, a fresh ansatz circuit is built with the current parameter values, the initial state of the circuit is reset to $|0\rangle^{\otimes n}$, the circuit is applied to that state, and the expectation value of the Hamiltonian is read out. Then, the classical optimizer proposes a new set of parameters based on the cost value. The specific characteristics of the two classical optimizers under study, COBYLA and Adam with SPSA gradient estimation, are described later in Section~\ref{sec:Optimizers}. The optimization loop terminates when the optimizer meets its internal tolerance, when the cost function has not improved over 200 iterations, or when a maximum number of iterations is reached.

When the optimization loop terminates, the ansatz parameters are fixed, and the circuit is no longer updated. To extract a concrete schedule from the resulting quantum state, we sample bitstrings through repeated measurements in the computational basis. From this distribution, we select the bitstring with the highest probability over $50{,}000$ shots as the candidate schedule for the batch.

Finally, the assignments are used to update the state that is propagated to the next batch: the remaining workload of every task is decreased by the hours assigned in the current batch, and the optimized parameters of the ansatz are stored as the warm-start candidate for the next batch. The complete schedule is the concatenation of the per-batch assignments across all batches in the planning horizon.

\subsection{Classical optimizers}\label{sec:Optimizers}
For every iteration of the training loop, a classical optimizer proposes a new set of parameters based on the cost evaluations performed by the quantum side. The optimizer is the second main design choice studied in this work. The ansatz defines the geometry of the quantum landscape, while the optimizer determines how the search moves through it. We compare two methods of navigating the landscape: one operates without any gradient information, while the other estimates a gradient at every step.

\textbf{COBYLA} (\textbf{C}onstrained \textbf{O}ptimization \textbf{BY} \textbf{L}inear \textbf{A}pproximation) is a derivative-free algorithm~\cite{powell2007view}. At every iteration, it maintains $m+1$ points in the $m$-dimensional parameter space, where $m = 2n(r+1)$, and constructs a linear approximation of the cost function from evaluations at those points. The first iteration includes the $m+1$ cost evaluations required to build this initial approximation, and each subsequent iteration typically requires a single additional evaluation. It then proposes a trial step within a region whose radius adapts to the agreement between predicted and actual cost decreases. This method requires no derivatives and is well suited to problems in which each cost evaluation is expensive. We use the \texttt{SciPy} implementation with $\texttt{rho\_beg} = 0.2$, $\texttt{tol} = 10^{-3}$, and a maximum of $5000$ iterations.

\textbf{Adam with SPSA gradient estimation} combines two ingredients. The first is the \textbf{Adam} optimizer~\cite{kingma2014adam}, which maintains running estimates of the first and second moments of the gradient and uses them to compute adaptive learning rates. The second is \textbf{SPSA} (\textbf{S}imultaneous \textbf{P}erturbation \textbf{S}tochastic \textbf{A}pproximation), which is used here to estimate the gradient without resorting to per-parameter finite differences~\cite{spall1998overview}. At every iteration, SPSA samples a random perturbation vector $\Delta$ with components drawn uniformly from $\{-1, +1\}$, evaluates the cost at the two points $\theta \pm c \cdot \Delta$, and constructs the gradient estimate from the difference between these two evaluations. We use Adam with a learning rate of $\eta_0 = 0.01$, momentum parameters $\beta_1=0.9$ and $\beta_2 = 0.999$, learning-rate decay $\eta(t)=\eta_0/(1+0.01\cdot t)$, and SPSA schedule $c(t) = c_0/(t+1)^\gamma$, where $c_0 = 0.1$, $\gamma =0.101$, and $t$ is the iteration index within the training loop. As with COBYLA, we set a maximum of $5000$ iterations, which corresponds to up to $10{,}000$ cost evaluations for Adam+SPSA and up to $5000+m$ for COBYLA.

SPSA estimates the gradient with only two cost evaluations per iteration. In contrast, the parameter-shift rule would require $2\cdot 2n(r+1)$ evaluations to estimate the full gradient analytically~\cite{schuld2019evaluating}. SPSA is noisier and may require more iterations to reach a comparable cost, but its lower per-iteration cost can compensate for this drawback. Therefore, using SPSA instead of the parameter-shift rule reduces the overall execution time of the VQA. However, as we show in Section~\ref{sec:Results}, the per-iteration cost measured in our simulations is comparable for both optimizers, so the difference in total execution time is driven by the number of iterations each optimizer performs.

\subsection{Simulation setup}\label{sec:HPC}
All circuits are simulated on a single Bull Qaptiva 802 appliance using its \textbf{M}atrix \textbf{P}roduct \textbf{S}tate (MPS) simulator, which is available in addition to the simulators in the publicly documented myQLM software stack~\cite{myqlm}. An MPS simulation is exact as long as the bond dimension is not truncated. In our ansatz, single-qubit rotations do not change the bond dimension, while each CNOT gate, whose operator Schmidt rank is two, can at most double it across the cut that it crosses. Because every entangling stage of both the linear and brick-wall topologies contains exactly one gate acting on each pair of neighboring qubits, the bond dimension after $r$ repetitions is bounded by $\chi \le 2^r$; thus, $\chi \le 8$ for $r=3$, while the entanglement-free reference remains at $\chi = 1$. We set the maximum bond dimension to $16$, so all simulations reported in this work are exact up to numerical precision. All entangling gates act on nearest neighbors, so no SWAP gates are required.

We ensure reproducibility by fixing the random seed of each run, so that the initial parameters and the SPSA perturbation vectors are drawn from deterministic streams of random numbers.

Each iteration of the training loop involves MPS evolution through the ansatz and evaluation of the expectation value of the Ising Hamiltonian, together with additional per-iteration contributions whose importance we discuss in Section~\ref{sec:Results}. Because the cost of an MPS simulation grows with the bond dimension, the entanglement-free reference is cheaper to simulate than the entangled topologies for the largest batches. Differences in per-iteration runtime between ansatz families therefore reflect, in part, the cost of the simulator rather than the complexity of the circuit.

%

\section{Optimization and Results} \label{sec:Results}
\subsection{Experimental setup}\label{sec:Setup}
This section compares the six ansatz-optimizer configurations introduced in Section~\ref{sec:Development}: the linear, brick-wall, and entanglement-free ansatz families, each trained with COBYLA and with Adam using SPSA gradient estimation. All configurations are run on the same MS-RCPSP instance under identical conditions: the QUBO formulation described in Section~\ref{sec:QUBO}, the same Lagrange multipliers, batch decomposition with $d_{win} = 1$, $r = 3$ ansatz repetitions, a maximum of 5000 iterations, $50{,}000$ sampling shots, the same stopping criteria, and the same simulation environment described in Section~\ref{sec:HPC}. All six start from the same initial parameter vector in the first batch, so the comparison is paired with respect to initialization.

The test instance models a workforce of $6$ workers with heterogeneous skill profiles assigned to $6$ tasks with different priority weights and deadlines over a planning horizon of $12$ days. Table~\ref{table:Instance} shows the exact configuration of the instance: the workers, their daily capacities, and their respective skill profiles, as well as the tasks, total workloads, deadlines, and priority weights. Although the total workforce capacity over the planning horizon exceeds the total workload, the heterogeneous skill profiles and tight deadlines force the algorithm to prioritize among tasks. This setup is intended to represent the operational regime in which the algorithm would be invoked.

Although a schedule meeting all six deadlines exists for this instance, the batch decomposition distributes the remaining workload of each task uniformly over the days left until its deadline and has no visibility of future batches, which makes such a schedule non-trivial to reach. Deadline compliance is therefore reported as a secondary indicator of schedule quality, while capacity feasibility remains the primary criterion.

For each of the six configurations, the VQA is executed as described in Section~\ref{sec:VQA}. The full run of each configuration consists of up to 12 sequential training loops, one per day of the planning horizon. The qubit count of each batch is determined by the number of binary variables in the corresponding QUBO instance, with the largest batches operating at 30 qubits in the circuit.

\begin{center}
\begin{minipage}{\columnwidth}
    \centering
    \resizebox{\columnwidth}{!}{%
    \begin{tabular}{c c c c c c | c c c c c c c c}
     $R_i$&$\quad$& $\text{C}_i$&$\quad$& Skills $\mathcal{S}_i$&$\quad$&$\quad$&$\text{J}_j$&$\quad$&$\text{W}_j$(h)&$\quad$&$d_j$&$\quad$&$\pi_j$\\
    \hline
    $\text{R}_1$&&8h&&$\text{J}_2$, $\text{J}_4$&&&$\text{J}_1$&&24&&4&&100\\
    $\text{R}_2$&&8h&&$\text{J}_4$, $\text{J}_3$&&&$\text{J}_2$&&24&&6&&60\\
    $\text{R}_3$&&8h&&$\text{J}_3$, $\text{J}_1$&&&$\text{J}_3$&&64&&8&&50\\
    $\text{R}_4$&&8h&&$\text{J}_1$, $\text{J}_5$&&&$\text{J}_4$&&144&&12&&10\\
    $\text{R}_5$&&4h&&$\text{J}_5$, $\text{J}_6$&&&$\text{J}_5$&&96&&12&&10\\
    $\text{R}_6$&&4h&&$\text{J}_2$              &&&$\text{J}_6$&&24&&12&&10\\
    
\end{tabular}%
    }
    \makeatletter
    \def\@captype{table}
    \makeatother
    \caption{Composition of the MS-RCPSP test instance: six workers $R_i$ with their daily capacities $C_i$ and skill profiles $\mathcal{S}_i$, and six tasks $J_j$ with their total workload $W_j$, deadline $d_j$, and priority weight $\pi_j$.}
    \label{table:Instance}
\end{minipage}
\end{center}

\subsection{Computational cost}\label{sec:cost}
We first examine the computational cost of running the VQA for each of the six configurations. Figure~\ref{fig:avg_time} shows the average wall-clock time per optimizer iteration as a function of the qubit count of each batch, measured on the environment described in Section~\ref{sec:HPC}. Each curve contains one point per distinct batch size, namely 13, 19, 24 and 30 qubits.

\begin{center}
\begin{minipage}{0.95\columnwidth}
    \centering
    \includegraphics[width=\linewidth]{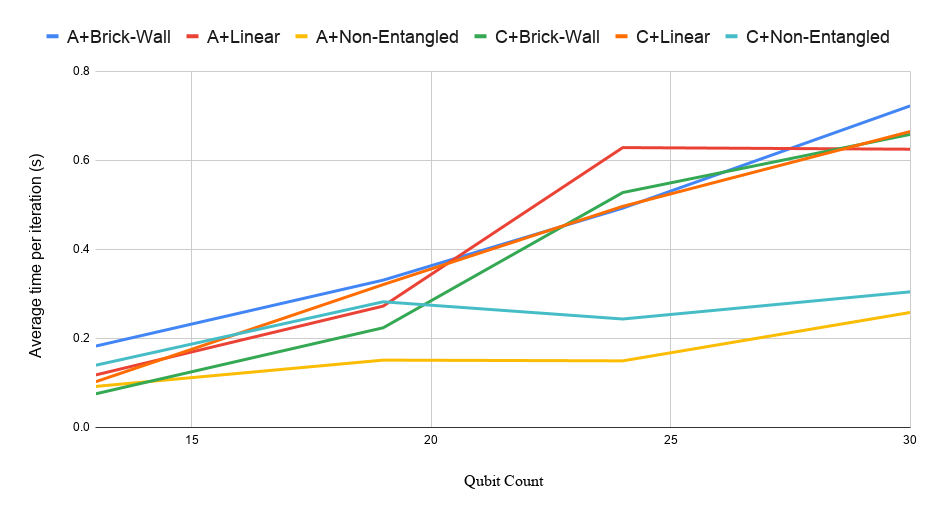}
    \makeatletter
    \def\@captype{figure}
    \makeatother
    \caption{Average wall-clock time per optimizer iteration as a function of the qubit count of the batch for the six ansatz-optimizer configurations. Each curve contains one point per distinct batch size. For COBYLA, the average includes the first iteration, which comprises the $m+1$ evaluations of the initial linear approximation. A: Adam+SPSA, C: COBYLA, Non-entangled: entanglement-free reference.}
    \label{fig:avg_time}
\end{minipage}
\end{center}

At the largest batch sizes, the entanglement-free reference is the cheapest ansatz family under both optimizers. At 30 qubits, its average time per iteration is approximately 0.26~s with Adam+SPSA and 0.30~s with COBYLA, compared with 0.63 to 0.72~s for the entangled topologies. At 13 and 19 qubits, however, the differences between families are small and not consistently ordered, and the entanglement-free reference is not always the cheapest configuration. This suggests that, for small batches, the per-iteration cost is dominated by contributions that do not depend on the entanglement of the state and that the growth of the bond dimension in the entangled topologies becomes the dominant contribution only for the largest batches. The linear and brick-wall topologies show no consistent ordering, which is consistent with both applying the same number of CNOT gates to the same qubit pairs.

Although each Adam+SPSA iteration requires two cost evaluations and each COBYLA iteration after the first only one, the per-iteration times of the two optimizers are comparable, and at 30 qubits Adam+SPSA is even slightly cheaper for the linear and entanglement-free families. This indicates that the measured time per iteration is not dominated by the cost evaluations alone, but also includes contributions paid once per iteration regardless of the optimizer. Some curves also show non-monotonic segments, such as the nearly constant cost of the linear topology with Adam+SPSA between 24 and 30 qubits.

The total execution time, which includes every training iteration and the final sampling step of each batch, also depends on the number of iterations performed, which differs markedly between configurations. With Adam+SPSA, the entangled topologies reach the 5000-iteration limit in most batches, whereas COBYLA stops before the limit in every batch. With COBYLA, the entanglement-free reference performs 8760 iterations over the full planning horizon, compared with 35{,}330 for the linear topology and 38{,}840 for the brick-wall topology, or roughly one-quarter as many. The linear topology with COBYLA completes all tasks by day 11, so no optimization is performed in its last batch. After the first three batches, the entanglement-free reference typically meets its tolerance after 110 to 350 iterations. In these batches, the trust-region radius reaches its lower bound after relatively few steps, suggesting that the starting point, which is usually inherited from the previous batch through the warm-start, is already close to a local minimum of the cost function.

As a result, the entanglement-free reference with COBYLA completes the full planning horizon in 2311~s, compared with 11{,}839~s for the linear topology and 11{,}210~s for the brick-wall topology, a reduction of approximately 80\% relative to both. For every ansatz family, COBYLA requires less total time than Adam+SPSA, a difference that arises mainly from the number of iterations each optimizer performs rather than from its cost per iteration.

These measurements characterize the cost of simulating the pipeline, not its cost on quantum hardware, where the relevant quantities are the number of circuit executions and the depth of the entangling stages.

\subsection{Training behavior}\label{sec:convergence}
We next examine how the six configurations behave during the training loop itself. We restrict this comparison to the first batch of the planning horizon, which is the only QUBO instance that is identical for all six configurations. This batch contains 30 qubits and, with $r = 3$ repetitions, produces an optimization landscape with 240 parameters. Figure~\ref{fig:cost_first_batch} shows the evolution of the cost function for the six configurations.

\begin{center}
\begin{minipage}{0.95\columnwidth}
    \centering
    \includegraphics[width=\linewidth]{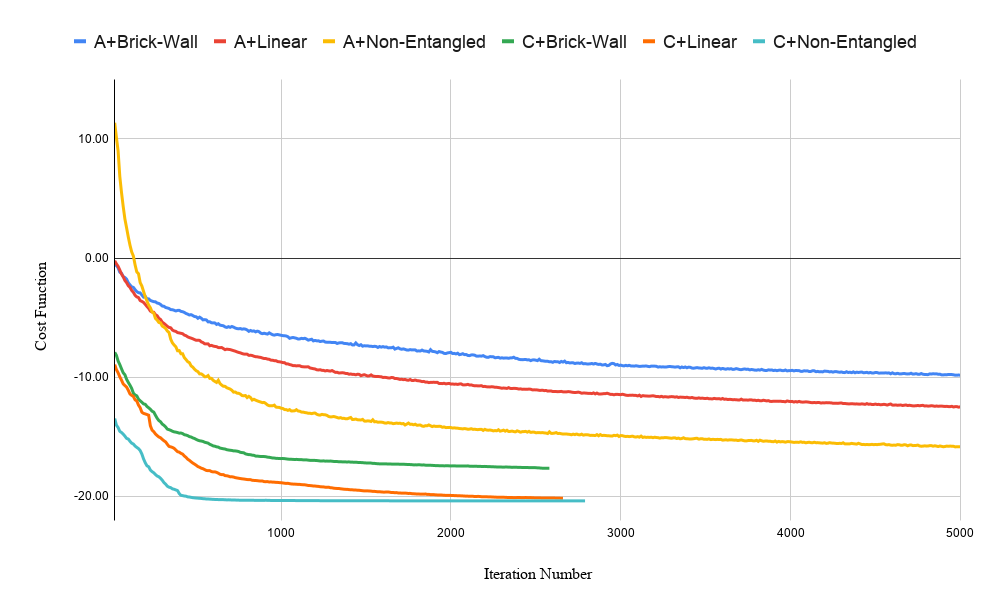}
    \makeatletter
    \def\@captype{figure}
    \makeatother
    \caption{Evolution of the cost function during the training loop of the first batch (30 qubits, 240 parameters) for the six ansatz-optimizer configurations. Each curve starts at the cost recorded after the first optimizer iteration. A: Adam+SPSA, C: COBYLA, Non-entangled: entanglement-free reference.}
    \label{fig:cost_first_batch}
\end{minipage}
\end{center}

Each curve starts at the cost recorded after the first iteration of the corresponding optimizer. Since the first COBYLA iteration already includes the construction of its initial linear approximation, whereas the first Adam+SPSA step changes each parameter by approximately the learning rate, the COBYLA curves start at lower cost values even though all configurations of the same ansatz family share the same initial parameters.

The trajectories separate first by optimizer. The three COBYLA configurations reach lower cost values than the three Adam+SPSA configurations and meet their tolerance after 2580 to 2790 iterations, whereas the Adam+SPSA trajectories are still slowly decreasing when they reach the 5000-iteration limit. Within each optimizer, the ordering of the ansatz families is the same: the entanglement-free reference reaches the lowest cost, followed by the linear topology and then the brick-wall topology. With COBYLA, the entanglement-free reference reaches its final cost within approximately 500 iterations and decreases only marginally afterward; the linear topology approaches a similar value only near the end of the optimization, and the brick-wall topology settles at a clearly higher cost. Note that the iteration axis corresponds to a different number of cost evaluations for each optimizer because each Adam+SPSA iteration requires two evaluations, whereas each COBYLA iteration after the first requires one.

A complementary view of the final state of each batch is provided by the probability of the selected bitstring, that is, the most probable bitstring in the final state. Table~\ref{table:probabilities} reports this probability for every batch, estimated from $50{,}000$ shots of the final MPS state. With COBYLA, the entanglement-free reference concentrates 98.4\% of the probability on the selected bitstring in the first batch and essentially 100\% in every subsequent batch, so its final state is, to a very good approximation, a single computational basis state. With Adam+SPSA, the entanglement-free reference reaches similar concentrations from the fifth batch onward, but not in the first four batches. In contrast, the entangled topologies remain spread over many bitstrings in most batches, with the selected bitstring carrying less than 1\% of the probability in the majority of cases and exceeding 50\% only in a few late batches of 13 qubits. For these configurations, the schedule is therefore extracted from a weakly concentrated distribution in which the selected bitstring is only weakly favored over many alternatives.

\begin{center}
\begin{minipage}{\columnwidth}
    \centering
    \resizebox{\columnwidth}{!}{%
    \begin{tabular}{c | c c c c c c}
     Batch & A+NE&A+L&A+BW&C+NE&C+L&C+BW\\
    \hline
    1&0.01\%&0.00\%&0.00\%&98.44\%&0.12\%&0.06\%\\
    2&0.16\%&0.01\%&0.00\%&99.98\%&0.01\%&0.10\%\\
    3&0.12\%&0.08\%&0.01\%&100.00\%&0.39\%&0.10\%\\
    4&4.29\%&0.16\%&0.01\%&100.00\%&0.78\%&0.09\%\\
    5&91.72\%&0.01\%&0.02\%&100.00\%&0.01\%&0.04\%\\
    6&99.88\%&0.02\%&0.12\%&100.00\%&0.02\%&5.91\%\\
    7&99.81\%&0.20\%&0.13\%&100.00\%&4.99\%&0.00\%\\
    8&99.45\%&0.20\%&0.14\%&100.00\%&8.00\%&0.00\%\\
    9&19.28\%&6.58\%&0.14\%&100.00\%&0.00\%&0.00\%\\
    10&98.82\%&0.00\%&0.14\%&100.00\%&0.01\%&99.44\%\\
    11&99.98\%&13.24\%&0.14\%&100.00\%&50.06\%&99.95\%\\
    12&99.99\%&91.18\%&0.10\%&100.00\%&N/A&99.91\%\\
    \end{tabular}%
    }
    \makeatletter
    \def\@captype{table}
    \makeatother
    \caption{Probability of the selected bitstring, that is, the most probable bitstring in the final state, for each batch and configuration, estimated from $50{,}000$ shots. Values are rounded to two decimal places. N/A: no optimization was performed because all tasks had been completed. A: Adam+SPSA, C: COBYLA, L: linear, BW: brick-wall, NE: non-entangled (entanglement-free reference).}
    \label{table:probabilities}
\end{minipage}
\end{center}

The cost reached in the first batch does not by itself determine the quality of the full-horizon schedule, which depends on the bitstrings selected in every batch and on how the remaining workload propagates between them. The operational quality of the resulting schedules is discussed in Section~\ref{sec:quality}.

\subsection{Solution quality}\label{sec:quality}
Finally, we examine the quality of the schedules produced by each configuration over the full planning horizon. Table~\ref{table:solution_quality} reports, for each configuration, the number of capacity violations (the number of worker-day pairs for which the assigned hours exceed the worker's daily capacity), the number of tasks completed within their deadlines, the ratio between the total hours assigned and the total workload of 376 hours, the total hours assigned, and the average number of task switches per worker per day. For reference, it also reports the total number of iterations and the total execution time discussed in Section~\ref{sec:cost}.

\begin{table*}[t]
\centering
\begin{tabular*}{0.95\textwidth}{@{\extracolsep{\fill}} l c c c c c c}
Metric & A+NE & A+BW & A+L & C+NE & C+BW & C+L\\
\hline
Total capacity violations & 0 & 0 & 1 & 0 & 2 & 4 \\
Tasks meeting their deadlines & 4/6 & 5/6 & 4/6 & 5/6 & 5/6 & 6/6\\
Assigned/required (\%) & 99.4 & 105.3 & 103.1 & 102.1 & 103.7 & 104.8\\
Hours assigned & 374 & 396 & 388 & 384 & 390 & 394\\
Task switches per worker per day & 0.13 & 0.11 & 0.15 & 0.13 & 0.11 & 0.20\\
Total number of iterations & 32520 & 52570 & 50810 & 8760 & 38840 & 35330\\
Total execution time (s) & 5930 & 13641 & 17589 & 2311 & 11210 & 11839\\

\end{tabular*}
\caption{Operational quality and computational cost of the schedules produced by each ansatz-optimizer configuration over the full planning horizon. A: Adam+SPSA, C: COBYLA, NE: non-entangled (entanglement-free reference), BW: brick-wall, L: linear. Capacity violations count only worker-day pairs in which the assigned hours exceed the daily capacity. An assigned-to-required ratio above 100\% indicates over-allocation and does not imply that every task is fully covered. Total execution time includes all training iterations and the final sampling step of each batch.}
\label{table:solution_quality}
\end{table*}

Three configurations produce schedules without capacity violations: the entanglement-free reference under both optimizers and the brick-wall topology with Adam+SPSA. The entanglement-free reference is therefore the only ansatz family that yields a feasible schedule independently of the optimizer. Among the entangled configurations, the linear topology with COBYLA is the only one that completes all six tasks within their deadlines, but it does so with four capacity violations, making its schedule infeasible in practice. Meeting every deadline in this case comes at the expense of exceeding the daily capacity of some workers, which illustrates why capacity feasibility, rather than deadline compliance, is adopted as the primary quality criterion.

The entangled configurations also show that the ordering observed in the first batch does not carry over to the full horizon. Although COBYLA reaches lower cost values than Adam+SPSA in the first batch for every ansatz family, the entangled COBYLA configurations produce more capacity violations than their Adam+SPSA counterparts: four and two, compared with one and zero. As discussed in Section~\ref{sec:convergence}, the final states of the entangled topologies are weakly concentrated in most batches, so their schedules depend on which bitstring happens to be the most probable in a broad distribution.

\begin{figure*}[ht!]
\centering
\includegraphics[width=0.82\textwidth,height=0.65\textheight,keepaspectratio]{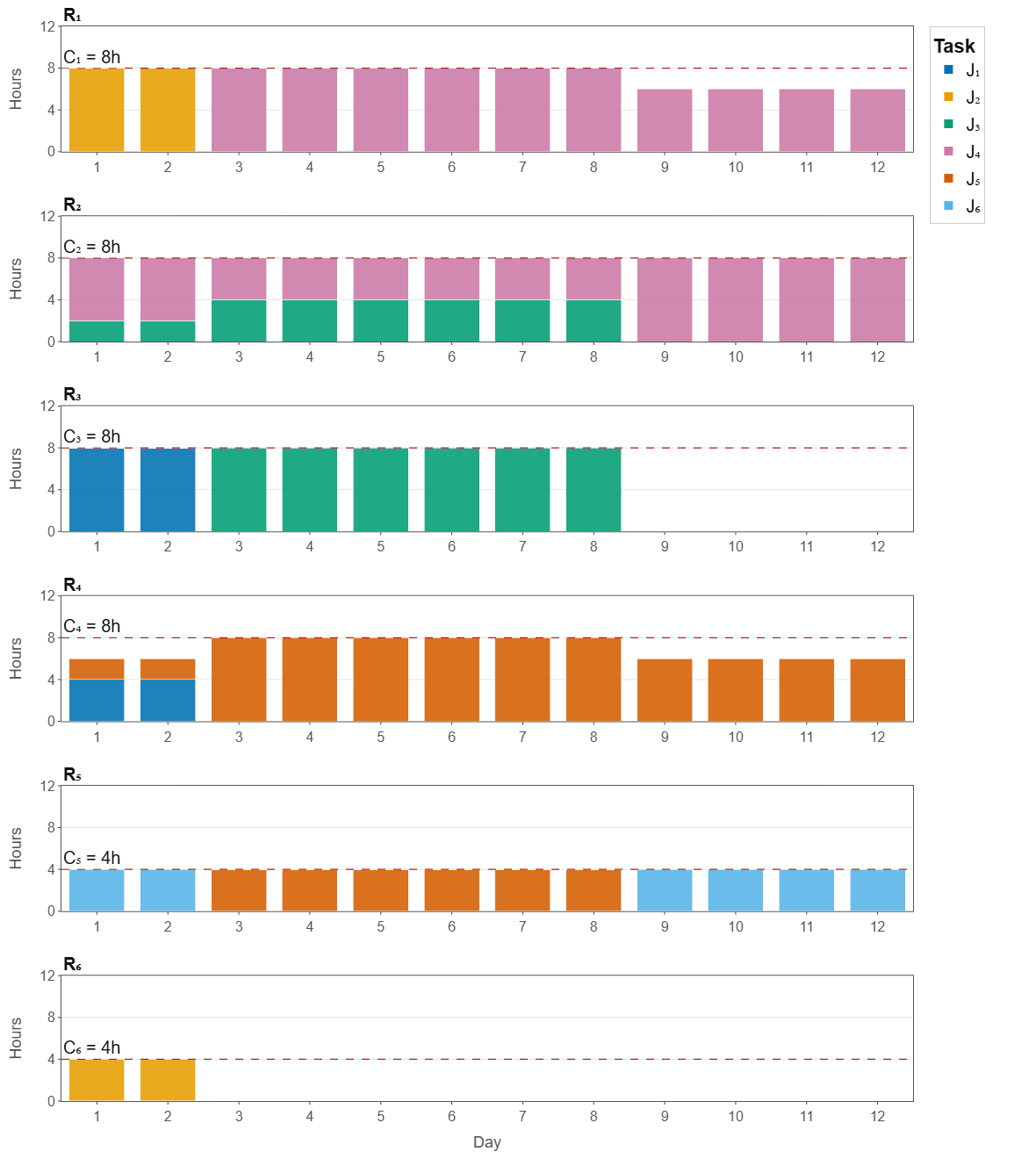}
\caption{Schedule produced by the entanglement-free reference with COBYLA for the test MS-RCPSP instance. Rows correspond to workers, colored segments indicate task assignments, and the dashed line marks the daily capacity.}
\label{fig:gannt}
\end{figure*}

Among the feasible schedules, the entanglement-free reference with COBYLA assigns 384 hours, or 102.1\% of the required workload, meets five of the six deadlines, and is obtained in 2311~s. The other two feasible configurations require 5930~s for the entanglement-free reference with Adam+SPSA and 13{,}641~s for the brick-wall topology with Adam+SPSA. Figure~\ref{fig:gannt} shows the schedule produced by the entanglement-free reference with COBYLA. Because hours can be over-assigned to some tasks while others remain partially uncovered, a ratio above 100\% indicates over-allocation rather than full completion. In this schedule, 2 hours of task $J_4$ remain uncovered.

These results correspond to a single shared initialization, so they describe the behavior of each configuration from a common starting point rather than its average performance. Confirming the ordering between configurations would require repeated runs and tests on different instances. Nevertheless, the concentration of the final states reported in Table~\ref{table:probabilities} indicates that the result obtained with the entanglement-free reference is the least sensitive to the sampling step because, with COBYLA, its final state is essentially a single bitstring in every batch after the first.


\newpage\hfill
\newpage\hfill
\newpage
\section{Conclusions}\label{sec:Conclusions}

\subsection{Interpretation of results}\label{sec:interpretation}
The central result of this work is that entanglement did not improve the schedules found by the VQA on the instance studied. The entanglement-free reference produced feasible schedules under both optimizers. With COBYLA, it produced the schedule closest to the required workload among the feasible ones, while requiring approximately 80\% less total execution time than the entangled topologies trained with the same optimizer. This is consistent with the fact that the ground state of the Ising Hamiltonian is a computational basis state, which a product state can represent exactly. However, this consistency does not by itself explain why the product state is also easier to train. A possible explanation is that the entangling stages add structure to the optimization landscape without adding representational power that the problem requires, but our results do not isolate this mechanism.

This result has direct implications for the evaluation of variational approaches to QUBO problems. For a product state, the expectation value of the Ising Hamiltonian reduces to a continuous relaxation of the QUBO that can be evaluated classically without simulating the circuit. The best-performing configuration in this work is therefore equivalent to a classical optimization of that relaxation. We consider this an informative outcome rather than a limitation of the study: it shows that, for hardware-efficient ansatz circuits of this depth, an entanglement-free reference with the same number of parameters is a necessary baseline and that any benefit attributed to entanglement in variational QUBO optimization should be demonstrated against it.

Regarding the optimizers, COBYLA reached lower cost values than Adam+SPSA in the first batch for every ansatz family and required less total execution time in every case. Because the measured cost per iteration is comparable for both optimizers, this difference arises mainly from the number of iterations: COBYLA meets its tolerance before the iteration limit, whereas Adam+SPSA exhausts the limit in most batches. However, the advantage of COBYLA in the first batch did not translate into more feasible schedules for the entangled topologies, whose COBYLA configurations produced more capacity violations than their Adam+SPSA counterparts. The entanglement-free reference was the only ansatz family whose feasibility was robust to the choice of optimizer.

These conclusions are subject to four limitations. First, they are based on a single instance and a single shared initialization, so they describe the behavior of each configuration from a common starting point rather than its average performance. Second, the entangled topologies considered here generate limited entanglement. With $r = 3$ nearest-neighbor entangling stages, the bond dimension is bounded by $8$, so our results do not rule out a benefit from deeper or problem-tailored entangling structures. Third, we did not compute the exact optimum of each batch, so the comparison is relative among configurations and does not quantify their distance from the optimal schedule. Finally, the batch decomposition fixes the schedule one day at a time, which constrains the quality attainable by any ansatz and may contribute to task $J_4$ remaining partially uncovered in the schedule obtained with the entanglement-free reference and COBYLA.

\subsection{Future work and conclusions}
Several directions follow from these observations. The first is to test whether the advantage of the entanglement-free reference holds across multiple initializations, on larger instances with denser couplings between variables, and with multi-day batches, in which correlations between variables on different days could make entanglement more relevant. The second is to compare the variational approach with a problem-tailored QAOA ansatz, quantum annealing formulations of the same QUBO, and classical solvers such as MILP and metaheuristics, together with the exact optimum of each batch where it is computable. The third concerns the optimization itself, including adaptive and natural-gradient optimizers and a systematic calibration of the Lagrange multipliers, which were set manually in this work.

This work has presented a VQA pipeline for the MS-RCPSP, including a QUBO formulation with worker capacities, skill profiles, task priorities, deadlines and an optional calendar-stability term, a logarithmic binary encoding, and a temporal batch decomposition that keeps every batch at or below 30 qubits. By comparing linear and brick-wall hardware-efficient ansatz circuits with an entanglement-free reference of equal parameter count, we found that, for the instance and initialization studied, entanglement increased simulation cost and spread the final state over many bitstrings without improving the resulting schedule, while the entanglement-free reference with COBYLA produced a feasible schedule at a fraction of the cost of the entangled configurations. These results indicate that the benefit of entanglement in variational QUBO optimization should not be assumed, and that entanglement-free baselines should be part of the standard evaluation of such algorithms.




\hbadness=10001
\bibliography{references}

\end{document}